# Crystal growth and characterization of the ultra-high temperature substrate $Ta_{1-x}Hf_xC_{0.5}$

Evan N. Crites[1,2,*], Sharad Mahatara[3], Joshua R. Hummel[4], Sydney R. Laywell[1], Ahamed Raihan[5], Shivashree S. Gowda[6], Ethan A. Scott[6], Amitayush Jha Thakur[7], Jessica L. McChesney[7], Patrick E. Hopkins[6,8,9], MVS Chandrashekhar[5], Michael G. Spencer[5], Stephan Lany[3], Satya K. Kushwaha[1,2], Tyrel M. McQueen[1,2,4,*]

Affiliations:

[1]*Department of Chemistry, The Johns Hopkins University, Baltimore, Maryland 21218, United States*

[2]*William H. Miller III Department of Physics and Astronomy, The Johns Hopkins University, Baltimore, Maryland 21218, United States*

[3]*National Laboratory of the Rockies, Golden, CO 80401, United States*

[4]*Department of Materials Science and Engineering, The Johns Hopkins University, Baltimore, Maryland 21218, United States*

[5]*Center for Research and Education in Microelectronics, Department of Electrical and Computer Engineering, Morgan State University, Baltimore, MD 21251, United States*

[6]*Department of Mechanical and Aerospace Engineering, University of Virginia, Charlottesville, Virginia, 22904, United States*

[7]*Advanced Photon Source, Argonne National Laboratory, Lemont, Illinois 60439, USA*

[8]*Department of Materials Science and Engineering, University of Virginia, Charlottesville, Virginia, 22904, United States*

[9]*Department of Physics, University of Virginia, Charlottesville, Virginia, 22904, United States*

**Abstract**

Incorporation of $Al_yGa_{1-y}N$ semiconductors into high power electronics offers efficiency improvements in power transmission, generation, and use, if approaches to eliminate the defects arising from film-lattice mismatch can be established. Here, we report the optical floating zone crystal growth of $Ta_{1-x}Hf_xC_{0.5}$ (x = 0.2), a new metallic substrate material family lattice matched to the ultra-wide-band-gap, Al-rich side (y = 0.91) of the $Al_yGa_{1-y}N$ solid solution. Laue diffraction demonstrates large single crystal domains in the as-grown boule. Single crystal x-ray diffraction at $T$ = 213 K in conjunction with first principles calculations shows that the material adopts a layered crystal structure with AA-type stacking of (Ta/Hf)-C-(Ta/Hf) trilayers described in the trigonal space group *P-3m1* (#164), with a = 3. 1168(4) Å, c = 4.9644(4) Å, and β = 120.0°. X-ray photoelectron spectroscopy (XPS) measurements show the Hf:Ta ratio to be close to the nominal value of 0.8:0.2 in the grown crystal. Density Functional Theory calculations reveal that this structure is stabilized by the low energy of carbon-vacancy formation of a hypothetical $(Ta/Hf)_1C_1$ anti-NiAs structure type, and imply flexibility in interface structure with an overlayer nitride film. A surface preparation/polishing procedure is developed that reduces root mean square (RMS) surface roughness from as-cut 130 nm to 7 nm as measured by atomic force microscopy. Scanning electron microscopy shows the presence of a native surface oxide, removed by polishing, along with carbon-rich pits. Time-domain thermoreflectance measurements show a room temperature thermal conductivity of $\kappa$ = 18.1(4) W $m^{-1}$ $K^{-1}$. These results provide key first steps for utilizing metallic, lattice matched, substrates for the growth of Al-rich $Al_yGa_{1-y}N$ semiconductors.

**Introduction**

The ultra-wide, direct, band-gap of $Al_yGa_{1-y}N$ semiconductors allow for a substantial decrease in the energy lost in energy conversion and transmission, greatly improving power efficiency, and such technology is now widely deployed for low-power applications such as such as radar,[1] consumer electronics,[2,3] and space applications.[4–6] Bringing these gains to high power applications, such as rapid electric vehicle charging[7,8] and highly efficient power distribution,[1,9] requires increasing voltage and current tolerance by reducing crack defects and grain boundaries, and moving to more Al-rich formulations.

Current state-of-the-art films are produced by growing a buffer layer of GaN on a sapphire or Si substrate. However, there will always be a mismatch between the grown material and a Si or sapphire substrate, a source of edge and screw dislocations. In sapphire this mismatch is ~13%,[10] while dislocation-free epitaxial matches generally require <2%.[11] $ZrB_2$ has been reported to lattice match GaN perfectly, however it lacks tunability in unit cell parameters.[12] While these other substrates are used due to their high availability, low cost, or excellent match to GaN, an ideal substrate would be metallic, to enable efficient heat dissipation and provide a back contact, and have a tunable chemistry to be lattice matched to a range of y-values in $Al_yGa_{1-y}N$. The basal plane lattice parameter of $Al_yGa_{1-y}N$ ranges from a = 3.11 Å in AlN (y = 1) to a = 3.19 Å in GaN (y = 0).[13] Closest packed ABC stacking (fcc cubic, <1 1 1> directions, a = 4.40-4.51 Å), AB stacking (hcp hexagonal, <0 0 1> directions, a = 3.11-3.19 Å), and AA stacking (trigonal, <0 0 1> directions, a = 3.11-3.19 Å) intermetallics are thus attractive choices, Figure 1.

Cubic TaC (a = 4.453 Å) and trigonal $Ta_2C$ (a = 3.106 Å) are known materials meeting these requirements for y = 0.52 and y = 1.0 respectively.[14,15] Initial synthesis efforts have demonstrated strain-free interfaces between sputtered TaC virtual substrates and AlGaN films grown by molecular epitaxy.[16] First principles calculations have determined the stable TaC/AlN and TaC/GaN interface atomic structures under variation of synthesis conditions.[17] Device simulations based on electronic structure calculations for these interfaces have predicted the possibility to achieve significant performance improvements with TaC/$Al_{0.5}Ga_{0.5}N$ heterostructures, highlighting the potential of such unconventional substrates for applications.[18]

These materials can be further tuned to produce substrate families that can be tailored for different Al/Ga ratios. HfC, a known compound isostructural to TaC, has a = 4.631 Å and, when made in a solid solution of $Ta_{1-x}Hf_xC$, would lattice match $Al_yGa_{1-y}N$ for y = 0-0.52. No $Hf_2C$ analog of $Ta_2C$ is known, but a hypothetical $Ta_{1-x}Hf_xC_{0.5}$ solid solution would lattice match y = 1 to 0.45 based on differences in atomic radii. Thus a combination of $(Ta_{1-x}Hf_x)C$ and $(Ta_{1-x}Hf_x)C_{0.5}$ would provide a single chemical system capable of lattice matching any $Al_yGa_{1-y}N$ film. Unfortunately, the high melting point of the $Ta_{1-x}Hf_xC_z$ family members has prevented its use as a single crystal substrate. For example, nowhere along the $Ta_{1-x}Hf_xC$ phase diagram is there a liquid below 3900 °C.[19] This high melting point has meant that single crystals of $Ta_{1-x}Hf_xC$, of any composition where x ≠ 1 or 0 have yet to be synthesized despite previous attempts dating back to the 1960s.[20–22] Organometallic[23] and laser melted[24,25] routes of crystal formation have been attempted, but yielded polycrystals. Further, there is no evidence in the literature of being able to produce single crystalline trigonal $Ta_{1-x}Hf_xC_{0.5}$ for x > 0.

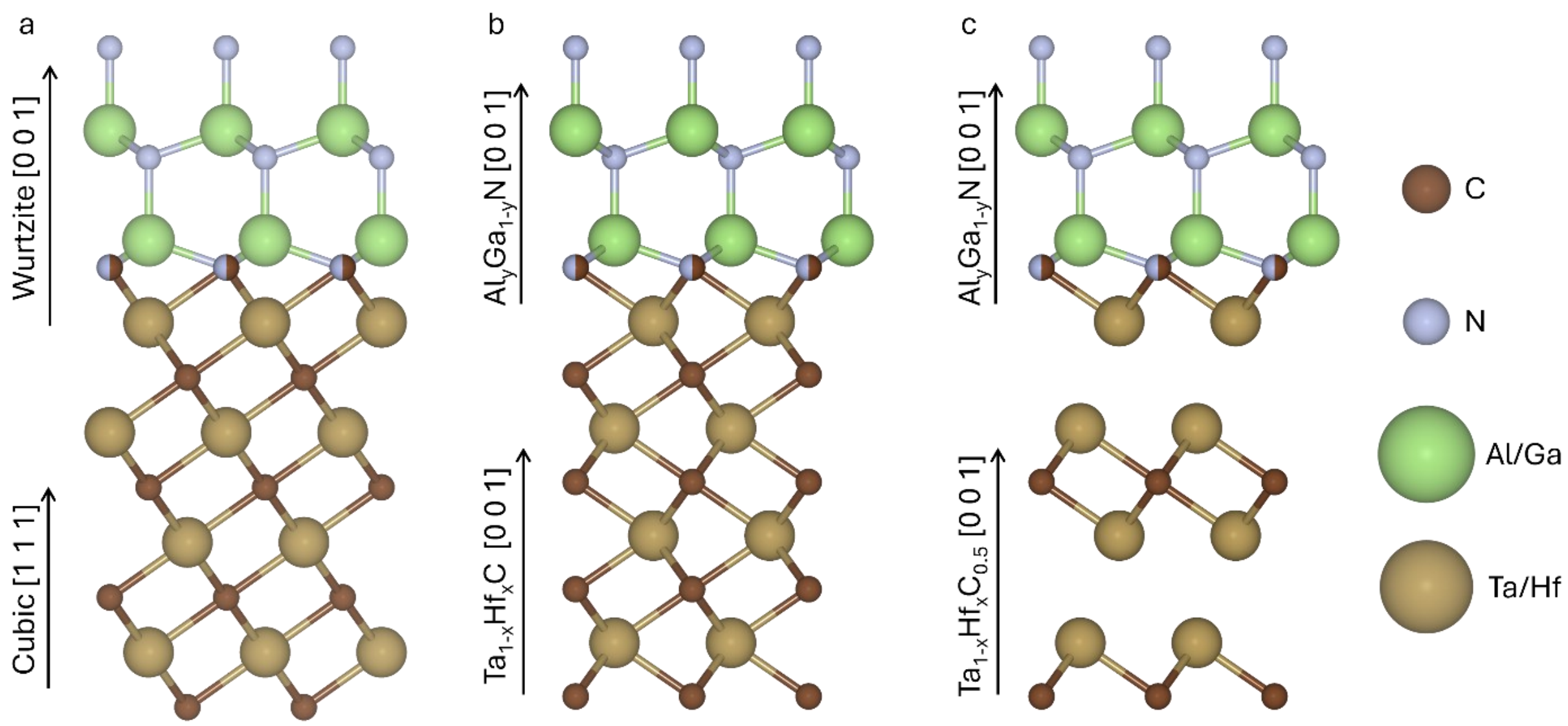

**Figure 1**. Schematic (a) fcc cubic-wurtzite, (b) hcp hexagonal-wurtzite, and (c) trigonal-wurtzite interfaces demonstrating their ability to serve as ideal substrates for $Al_yGa_{1-y}N$. In this work, we report the development of a new trigonal substrate family as drawn in (c).

Here, we report the growth, crystal structure solution, characterization, and preparation of a single crystal of AA stacking (trigonal) $Ta_{0.8}Hf_{0.2}C_{0.5}$ for use as an $Al_yGa_{1-y}N$ growth substrate. Single crystal growth starting from polycrystalline Ta-Hf-C rods produced by arc melting was performed in a specially-tailed laser floating zone furnace designed for ultra-high temperature operation ($T > 3500$ °C). Laue diffraction shows large, orientable single crystal regions. First principles calculations were used to identify plausible structure types with different C stoichiometry. Single crystal x-ray diffraction (SCXRD) was used to solve the structure, a layered crystal structure with AA-type stacking of (Ta/Hf)-C-(Ta/Hf) trilayers described in the trigonal space group *P-3m1* (#164), with a = 3.1168(4) Å, c = 4.9644(4) Å, and β = 120.0° at $T$ = 213 K, in good agreement with the computational results. The calculations also reveal that this structure is stabilized by the low energy of carbon-vacancy formation of a hypothetical $(Ta/Hf)_1C_1$ anti-NiAs structure type, and imply flexibility in interface structure with an overlayer nitride film. A mechanical polishing procedure reduces root-mean square (RMS) surface roughness from as-cut 130 nm to 7 nm as measured by atomic force microscopy (AFM) and prepares the surface for planarization and cleaning. Scanning electron microscopy (SEM) was used to directly compare a polished and unpolished crystal segment and shows the presence of a native surface oxide that is removed by polishing, along with carbon-rich inclusions. X-ray photoelectron spectroscopy (XPS) measurements help to further quantify the Hf:Ta ratio of a cleaved surface. Frequency-dependent time-domain thermoreflectance (TDTR) measurements yield a room temperature thermal conductivity of $\kappa = 18.1(4)$ W $m^{-1}$ $K^{-1}$. These results provide key first steps for utilizing metallic, lattice matched, substrates for the growth of Al-rich $Al_yGa_{1-y}N$ semiconductors.

**Methods**

Carbon rods (Alfa Aesar, 99.9995% metals basis, lot #L17M14) were ground into chunks or lathed into powder for use. Ta pellets (Kurt J. Lesker, 99.95% pure, lot #TAF2TA3654727A031122/VR10-0053090) and Hf pellets (Kurt J. Lesker, 99.9% pure excluding Zr, lot #HFF415199030521/VR10-0032926) were used without further purification. Metal pellets of Ta and Hf were combined in a 4:1 Ta:Hf molar ratio with a 110% excess of C in powder and chunk form. To mix and react the material, an arc melter was used with a maximum current of 250 A under partial vacuum of an inert Ar atmosphere. Ta and Hf were added in a stoichiometric ratio totaling 14.4078 g (79.62 mmol) of Ta and 3.3534 g (18.79 mmol) of Hf. A total of 1.2949 g (107.9 mmol) of C was slowly added and the billet was melted until two criteria were met to signify C incorporation: (1) the structure changing as measured by powder x-ray diffraction and (2) the material requiring a higher current to melt, signifying the composition of the melt had changed enough to cause a higher melting point. At that point the bulky rod was cut down its length and further melted to shape into a longer and thinner rod for growth.

To prepare the rod for floating zone (FZ) growth, it had to be shaped into a uniform cross-section. $Ta_{0.8}Hf_{0.2}C_{0.5}$ is a very hard material (~30 GPa)[26] and cutting it is difficult, even with diamond-embedded wires and blades. The precise cuts required for preparation for FZ growth are even more difficult. A Fanuc α-oiE with a 0.01 inch diameter wire was used for wire electro discharge machining (EDM) to shape the rod into a rectangular bar so it would have a uniform cross-section for consistent mass flow into and out of the molten zone. The layer deposited on the rod from the EDM was removed before growth using SiC polishing pads. The growth was conducted in a custom-built laser FZ operating in the 6-laser configuration, allowing for a maximum of 2400 W of power under flowing Ar (Figure S1).[27] The rods were initially rotated at 6 rpm in the same direction during formation of the molten zone and joining of the rods. Melting initially occurred at ~35% power, with growth taking place at 50% laser power. The rods were counter rotated during growth, with the top being rotated clockwise and the bottom being rotated counterclockwise, both at 5 rpm and translated at 10 mm/hr for ~20 minutes followed by 5 mm/hr for 3.25 hour to yield a rod approximately 18 mm long.

Laue was obtained along the grown rod using a Multiwire Laboratories Ltd. MWL120 Real-Time Back-Reflection Laue Camera System controlled with a MWL 731 Six-Axis Motor Driver with a Spellman DF60N3 x-ray Generator providing an x-ray source. The generator was run at 10 mA, between 10 and 15 kV, with 15 kV giving better contrast. Spots demonstrated the crystallinity of a region in the center where single crystal samples were extracted from. The polycrystalline lower portion of the rod was used to trial the polishing steps. Polishing was done by affixing pieces to a Buehler Ecomet 30 rotary mechanical polisher using CrystalBond. The CrystalBond was not strong enough to stand up to the pressures due to the forces delivered by the polisher over the small surface area of the pieces. A holder for the pieces was designed using an aluminum disk with appropriately sized holes drilled into it. This further helped to slow down the grinding and allowed for more careful material removal, as the aluminum disk took the brunt of the force while also having the same thickness as the sample pieces to allow for a flat face. The grit number of the polishing pads was increased in a stepwise manner, moving to higher grits and smaller particle sizes to achieve a mirror-like finish on the cut pieces. Initially, 120 grit SiC sandpaper was used to eat through the aluminum plate holding the pieces in place and grind away the sample until the pieces were flat. Polishing then proceeded using 240, 400, 600, 800, and 1200 grit SiC sandpaper in that order. Finishing mechanical steps were done using 1, 0.5, and 0.1 µm particle size diamond polishing sheets obtained from UltraTec. An Olympus BX51 Microscope was used to take images of the polished surfaces between steps.

The polished surfaces were imaged using a Zeiss Evo 10 Scanning Electron Microscope (SEM) with EDAX SmartEDX Energy-Dispersive X-Ray Spectroscopy Analysis System, 30 mm$^2$, 129 eV, 600 kcps, Silicon Nitride Window. A Dimension 3100 Nanoscope IV in tapping mode with Pt coated tips ~60 kHz resonance frequency atomic force microscope was used to characterize surface roughness. Single-crystal X-ray diffraction data was collected at $T$ = 213 K using a SuperNova diffractometer (equipped with an Atlas detector) with Mo Kα radiation ($\lambda$ = 0.71073 Å) (Rigaku Synergy DW). Data reduction and absorption correction were performed using the CrysAlis PRO software.[28] Refinements of the structural models were performed using the SHELX software operating via WinGX.[29,30] Visualizations of the models were performed using the VESTA software.[31]

Thermal properties of a polished $Ta_{0.8}Hf_{0.2}C_{0.5}$ disk were measured using the frequency-dependent time-domain thermo-reflectance (TDTR) technique with a two-tint configuration.[32,33] A general schematic of a TDTR setup is shown in Figure S2. A Ti:sapphire laser with a wavelength of ~800 nm at 80 MHz is used for pulse generation. The laser output is split into a pump and a probe beam by a polarized beam splitter (PBS). The 5.5 ns of delay between pump and probe pulses is achieved by a linear motorized stage in the pathway of the probe beam. Modulation of the pump beam is achieved by an electro-optic modulator (EOM) operating over a frequency range of 0.1-10 MHz. The pump and probe paths are collinearized with a PBS and concentrically focused on the sample with a 10X objective lens, providing $1/e^2$ pump and probe diameters of ~12 µm. An Al transducer deposited on the $Ta_{0.8}Hf_{0.2}C_{0.5}$ disk is heated to a temperature increase of < 10 K by the modulated pump beam. The probe beam measures the change in reflectance at the surface due to the change in temperature. A ThorLabs Model PDB415A balanced photodetector coupled to a Zurich Instruments UHFLI digital lock-in amplifier monitors the probe beam. The lock-in amplifier filters the probe beam signal and outputs the in-phase and out-of-phase components. Extraction of thermal properties is accomplished by fitting the negative ratio of the in-phase and out-of-phase components to the thermal model.[34]

X-ray photoemission spectroscopy (XPS) measurements were performed at beamline 29-ID of the Advanced Photon Source, Argonne National Laboratory. Synchrotron radiation with a photon energy of 1486 eV was used as the excitation source. The emitted photoelectrons were analyzed using a Scienta R4000 hemispherical analyzer with a pass energy of 200 eV (with an experimental energy resolution of 200 meV, combined X-ray and analyzer). The sample temperature was maintained at < 80 K during the measurements. The binding energies of core levels were calibrated with respect to the Fermi level. The samples were annealed in vacuum at 900 C for 10 mins before measurement. The C:Ta ratio was estimated from peak ratios extracted from Voigt fits after removing the Shirley background, to C 1*s* carbide peak located at 282.9 eV and Ta 4*f* carbide doublet peaks located around 25 eV binding energies. The total Ta to Hf ratio was estimated from the total peak areas for Ta and Hf 4*f* peaks located around 25 eV and 17.5 eV binding energy respectively.

First principles calculations were performed with the projector augmented wave (PAW) method implemented in the VASP code,[35,36] using the Strongly Constrained and Appropriately Normed (SCAN) meta-GGA functional,[37] which affords a better description of lattice parameters and thermochemical properties[38] than standard LDA or GGA density functional theory (DFT) functionals. Per thermochemical convention, the elemental reference is taken as the phase that is thermodynamically stable at standard conditions, i.e. bcc Ta, hexagonal closed packed (hcp) Hf and graphitic carbon. A plane-wave cut-off of 380 eV is sufficient to yield converged total energies for the "Ta_pv", "Hf_pv", and "C_s" PAW potentials.[39] For the alloy calculations, we constructed a 2×2×2 supercell of the rock-salt conventional unit cell, containing 64 atoms, while for $Ta_2C$ structure, we expanded the hexagonal primitive cell into a (3,1)x(1,3) in-plane supercell and doubled it along the c-axis, yielding a 48-atom supercell. In these supercells, Ta sites were randomly substituted with Hf atoms up to 50% doping in steps of 6.25%. For each composition, three random alloy configurations were considered, and the respective alloy energy was taken as the average of these structures. A 4×4×4 k-point grid was used for the alloy supercell calculations, whereas all other calculations used Brillouin zone sampling of 5000 k-points per reciprocal atom. All structures were fully relaxed until the atomic forces were below 0.01 meV/Å.

**Results and Discussion**

Laue diffraction patterns (Figure 2) taken along the grown rod demonstrate an increase in crystallinity along the rod, with a sizable single domain segment approximately 1 cm from the start of the growth, and the trigonal/hexagonal direction perpendicular to the growth direction. The formation of two distinct domains along the crystal is due to two sizable domains not being pinched off via Ostwald ripening. A growth seeded on a single crystal with a specific orientation is expected to allow for the formation of a single crystalline domain.

SCXRD was used to determine the crystal structure. The data could be indexed equally well with two distinct but metrohedrally equivalent structures – an fcc cubic cell (a = 4.407 Å) and a trigonal/hexagonal cell (a = 3.1168(4) Å, c = 4.9644(4) Å, β = 120°). Preliminary refinements in these two choices yielded $R_1/wR_2$ = 4.15/12.23% for the former, and $R_1/wR_2$ = 3.2/7.71% for the latter. Combined with DFT analysis (*vida infra*), we thus select the trigonal/hexagonal cell. Figure 3(a-f) shows representative slices through the scattering volume. The pattern of systematic absences is most consistent with the trigonal space group P-3m1 (#164). The Ta:Hf ratio was fixed to the nominal ratio. A full structure refinement readily converges, and final parameters are given in Table 1, Table 2, and Table 3, and the structure visualized in Figure 3(g).

The finding of a trigonal structure is further supported by first-principles calculations given a carbon-deficient stoichiometry. The following structures were considered: (1) The cubic rocksalt (rs) structure (*Fm-3m*, #225), which is generally considered to be the ground state of TaC and HfC. (2) The NiAs structure (*P6₃/mmc*, #194), from earlier reporting for $Ta_{0.75}Hf_{0.25}C$ alloys,[21] with the metal on the 2*a* site and C on the 2*c* site. (3) The anti-NiAs structure in which the occupation is inverted, i.e., the metal on the 2*c* site and C on the 2*a* site. (4) The known trigonal structure of the carbon deficient $Ta_2C$ phase (*P-3m1*, #164).[15] Here, the metal is on the 2*d* site and C on the 1*a* site. Notably, this structure is related to the anti-NiAs structure by removal of one carbon layer. Thus, this C-deficient carbide phase with 2:1 composition provides a plausible candidate structure for the hexagonal phase.

The results of our SCAN calculations show that the rs structure is the ground state for both TaC and HfC, consistent with the previous report.[17] The NiAs phase is very unstable at 673 meV/f.u. (TaC) and 1101 meV/f.u. (HfC) above rs. The anti-NiAs phase lies lower in energy, but still considerably above rs at 214 meV/f.u. (TaC) and 436 meV/f.u. (HfC). Table I shows the calculated lattice parameters for the four considered crystal structures. For $Ta_2C$, the cell-internal Wyckoff parameter *z* for the Ta (2*d*) site is also given. After determining the lattice parameters of the binary end members, we estimated the structural parameters for the 80/20 Ta/Hf alloy using Vegard's-law interpolation (last column of Table 4). Because a 2:1 Hf-C phase has not been experimentally reported, we include it here only as a hypothetical analogue to $Ta_2C$.

The refinement results are in excellent agreement with the predicted 2:1 $Ta_{0.8}Hf_{0.2}C$ phase (sg 164) in Table I. The slight ~0.4% overestimation of the lattice parameters is within the typical accuracy of the SCAN functional.[37] In contrast, the anti-NiAs 1:1 structure (sg-194) reproduces the in-plane lattice constant but overestimates the c parameter by ~7%, far exceeding the expected SCAN error. This large discrepancy arises because the 2:1 structure is intrinsically more compressed along the c-direction, reflecting the removal of one carbon layer per formula unit relative to the 1:1 structure. Taken together, these results clearly indicate that the experimental crystal corresponds to a 2:1 (Ta,Hf)-C composition rather than a 1:1 phase.

**Table 1.** Refinement and lattice parameters calculated from Mo $K_\alpha$ x-ray data.

| Chemical Formula | $Ta_{0.8}Hf_{0.2}C_{0.5}$ |
|---|---|
| Crystal system | Hexagonal |
| Space Group | *P-3m1* (164) |
| Formula weight (g/mol) | 186.46 |
| Z | 2 |
| a (Å) | 3.1168(4) |
| c (Å) | 4.9644(4) |
| γ (°) | 120 |
| Volume (Å$^3$) | 41.7652(114) |
| $\rho_{calc}$ (g*cm$^{-3}$) | 14.827 |
| Mo $K_\alpha$ wavelength (Å) | 0.71073 |
| *T* (K) | 213 |
| $R_1$ (%) | 3.09 |
| $wR_2$(%) | 7.71 |
| GOF | 1.377 |

The relative stability of the 1:1 and 2:1 phases, i.e., $Ta_{1-x}Hf_xC$ and $Ta_{1-x}Hf_xC_{0.5}$, depends on the growth conditions. Results for the phase stability as function of the composition x and the carbon chemical potential $\Delta\mu_C$ is shown in Figure 4. The DFT-SCAN calculations (Figure 4) shows that 1:1 phase of TaC requires high carbon activity, $\Delta\mu_C = -0.5$ eV, which is difficult to maintain at high *T*.

Hf alloying progressively lowers the required $\Delta\mu_C$ enabling stabilization of the 1:1 phase under less carbon-rich conditions. Figure 4 shows the nearly linear decrease in $\Delta\mu_C$ as a function of Hf alloying. We also find that the hypothetical $Hf_2C$ phase is unstable with respect to HfC+Hf(s), consistent with the absence of an experimentally reported 2:1 hexagonal $Hf_2C$ phase.

**Table 2.** Atomic positions, occupancies, sites, and equivalent thermal displacement parameters. Ta1 and Hf1 were set to occupy a single site with their position and thermal displacement equal. Their occupancies were set to a 4:1 Ta:Hf ratio.

| Element | Label | Position (x, y, z) | SOF | Site | $U_{eq}$ |
|---|---|---|---|---|---|
| Ta | Ta1 | 1/3, 2/3, 0.25321(19) | 0.8 | 2d | 0.0062(4) |
| Hf | Hf1 | | 0.2 | | |
| C | C1 | 0, 0, 0 | 1 | 1a | 0.0014(50) |

**Table 3.** Anisotropic thermal displacement parameters. Ta1 and Hf1 were set to have their thermal displacement parameters equal, as they occupy the same site.

| Element | Label | $U_{11}$ | $U_{22}$ | $U_{33}$ | $U_{23}$ | $U_{13}$ | $U_{12}$ |
|---|---|---|---|---|---|---|---|
| Ta | Ta1 | 0.0054(4) | 0.0054(4) | 0.0074(5) | 0 | 0 | 0.0027(2) |
| Hf | Hf1 | | | | | | |

**Table 4.** SCAN calculated lattice parameters for the 4 considered crystal structures of Ta and Hf carbides in the 1:1 and 2:1 stoichiometry (stoi). For the cubic rs structure, the lattice parameters are given for the primitive cell. The conventional cell parameter equals a*√2.

| stoi | sg # | type | | *M*=Ta | *M*=Hf | *M*=$Ta_{0.8}Hf_{0.2}$ |
|---|---|---|---|---|---|---|
| *M*C | 221 | rocksalt | *a* (Å) | 3.143 | 3.260 | 3.166 |
| *M*C | 194 | NiAs | *a* (Å) | 3.125 | 3.240 | 3.148 |
| | | | *c* (Å) | 5.477 | 5.699 | 5.522 |
| *M*C | 194 | anti-NiAs | *a* (Å) | 3.081 | 3.298 | 3.124 |
| | | | *c* (Å) | 5.335 | 5.181 | 5.304 |
| $M_2C$ | 164 | $CdI_2$(hP3) | *a* (Å) | 3.096 | 3.253 | 3.128 |
| | | | *c* (Å) | 4.913 | 5.267 | 4.983 |
| | | | *z* | 0.2542 | 0.2368 | 0.2508 |
| Experimental (this work) | | | *a* (Å) | | | 3.1168(4) |
| | | | *c* (Å) | | | 4.9644(4) |
| | | | *z* | | | 0.2532(2) |

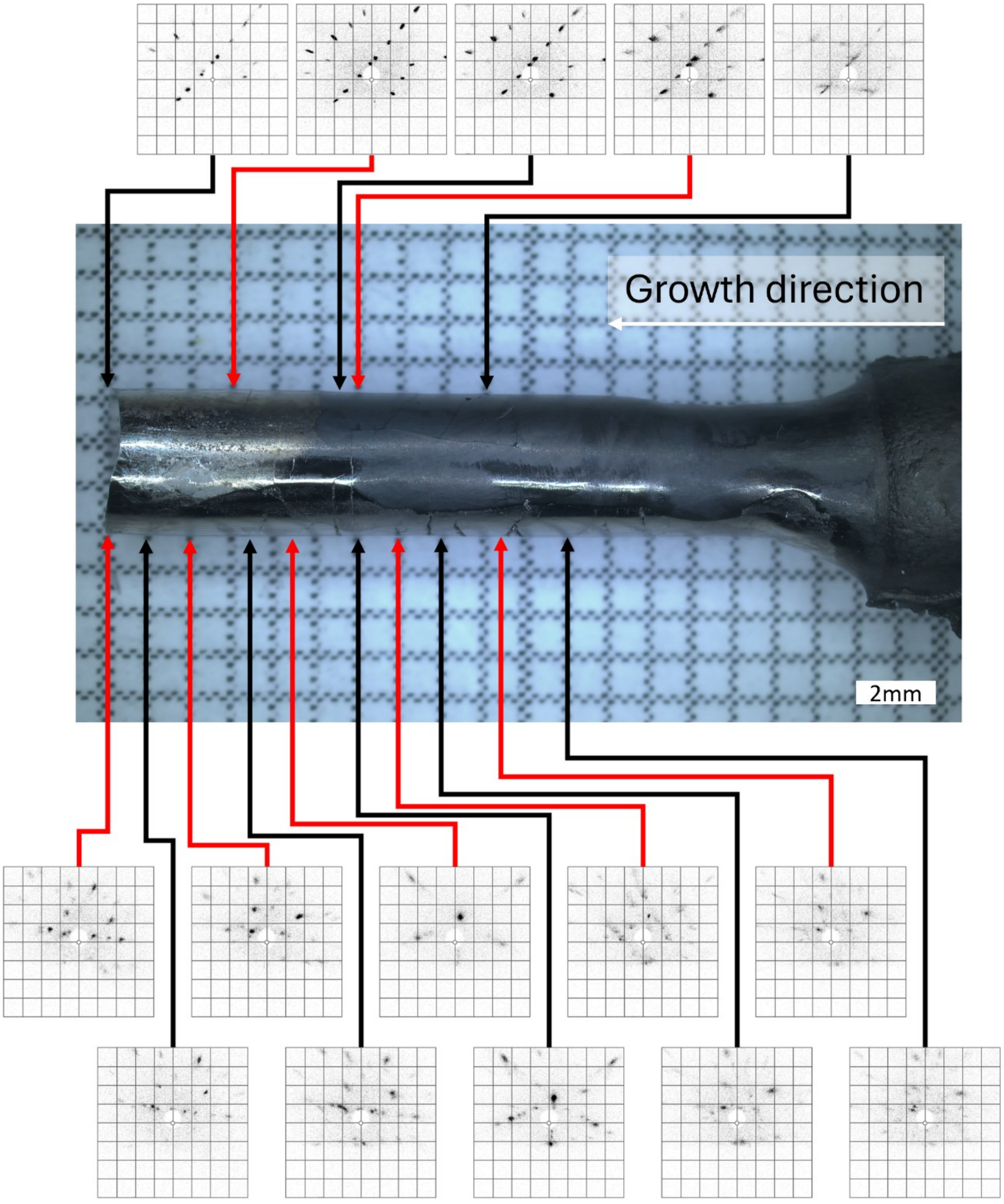

**Figure 2**. Optical image and Laue patterns along the grown rod showing increased crystallinity after approximately 1 cm of growth. One side of the rod was polished with SiC polishing pads, causing the diffraction patterns to be fuzzier due to the rough surface. In this image the polished side is the lower face. The unpolished side has much sharper spots, indicating the grown rod is made of two domains. In this image the unpolished side is the upper face.

The calculations also give insight into the chemical (dis)order present in $Ta_{1-x}Hf_xC_{0.5}$. When viewed as a carbon-vacancy ordered derivative of anti-NiAs $Ta_{1-x}Hf_xC_{1-\delta}$, a natural question is whether the high crystal growth temperature promotes significant disordering of C and vacant sites between layers. Taking $\Delta H(Va_C)$ = 0.31 eV, and a crystallization temperature of T = 3900 °C = 4173 K, and considering only the increase in configurational entropy from disorder, a simple min($\Delta H$-$T\Delta S$) calculation gives an estimated 20% C/vacancy mixing at the crystallization temperature. The presence of well-ordered C/vacancies thus suggests significant carbon mobility below the melting point, and implies that addition of a more gradual temperature gradient during cooling after crystallization could further help crystallographic perfection of this substrate material.

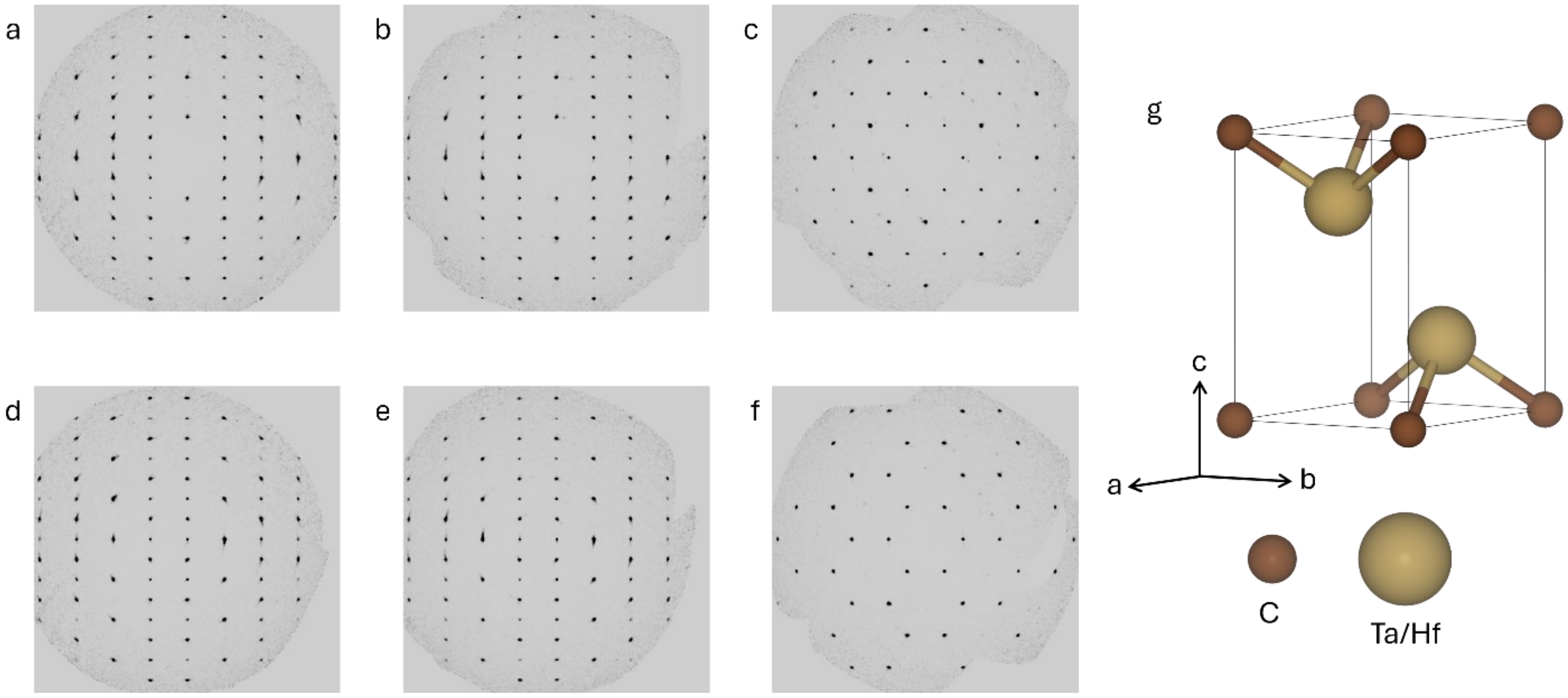

**Figure 3**. Single crystal x-ray diffraction in the (a) (0kl), (b) (h0l), (c) (hk0), (d) (1kl), (e) (h1l), and (f) (hk1) plane. (g) Solved crystal structure of $Ta_{0.8}Hf_{0.2}C_{0.5}$.

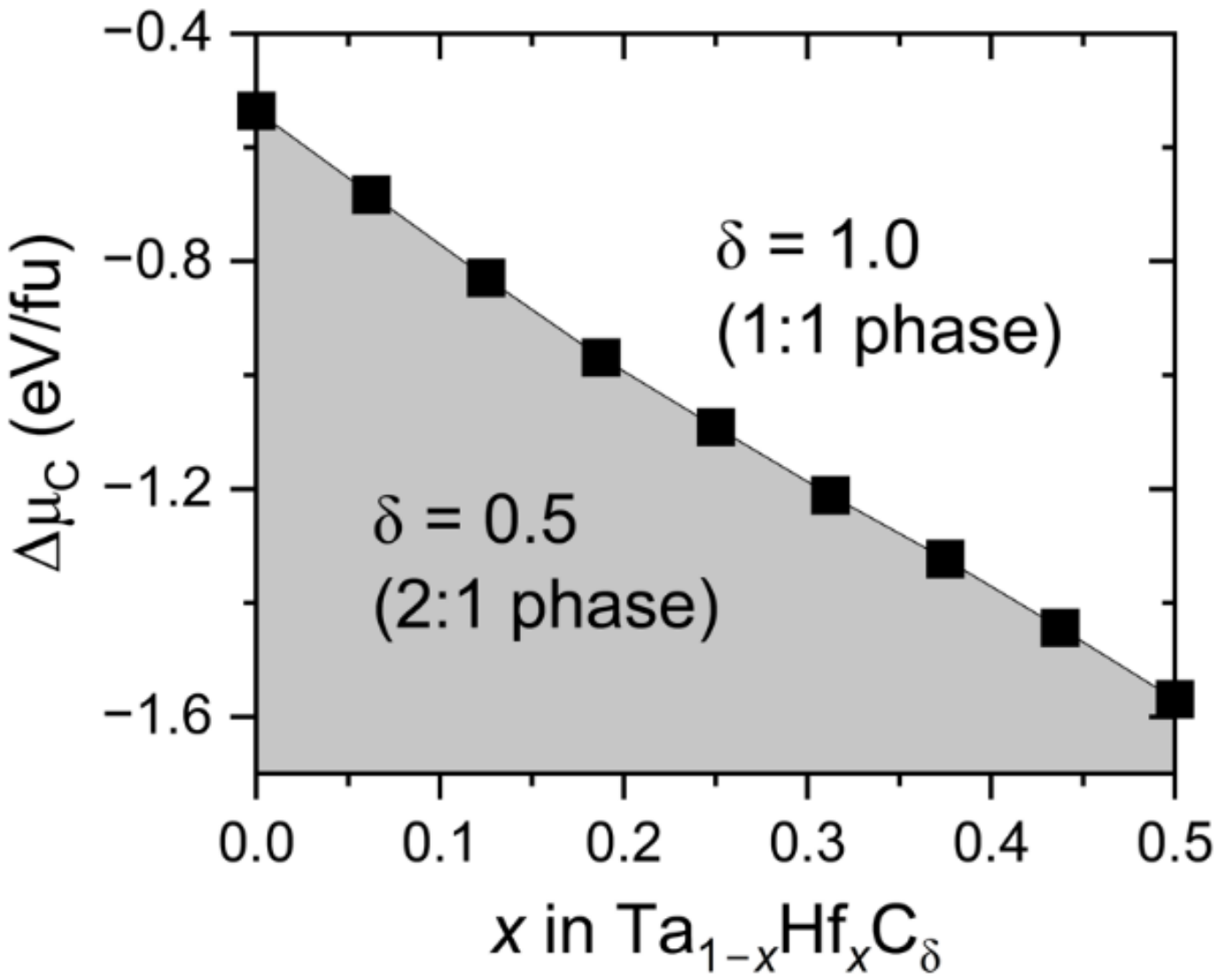

**Figure 4.** SCAN DFT computed phase map of phase stability as a function of carbon chemical position and Ta:Hf ratio.

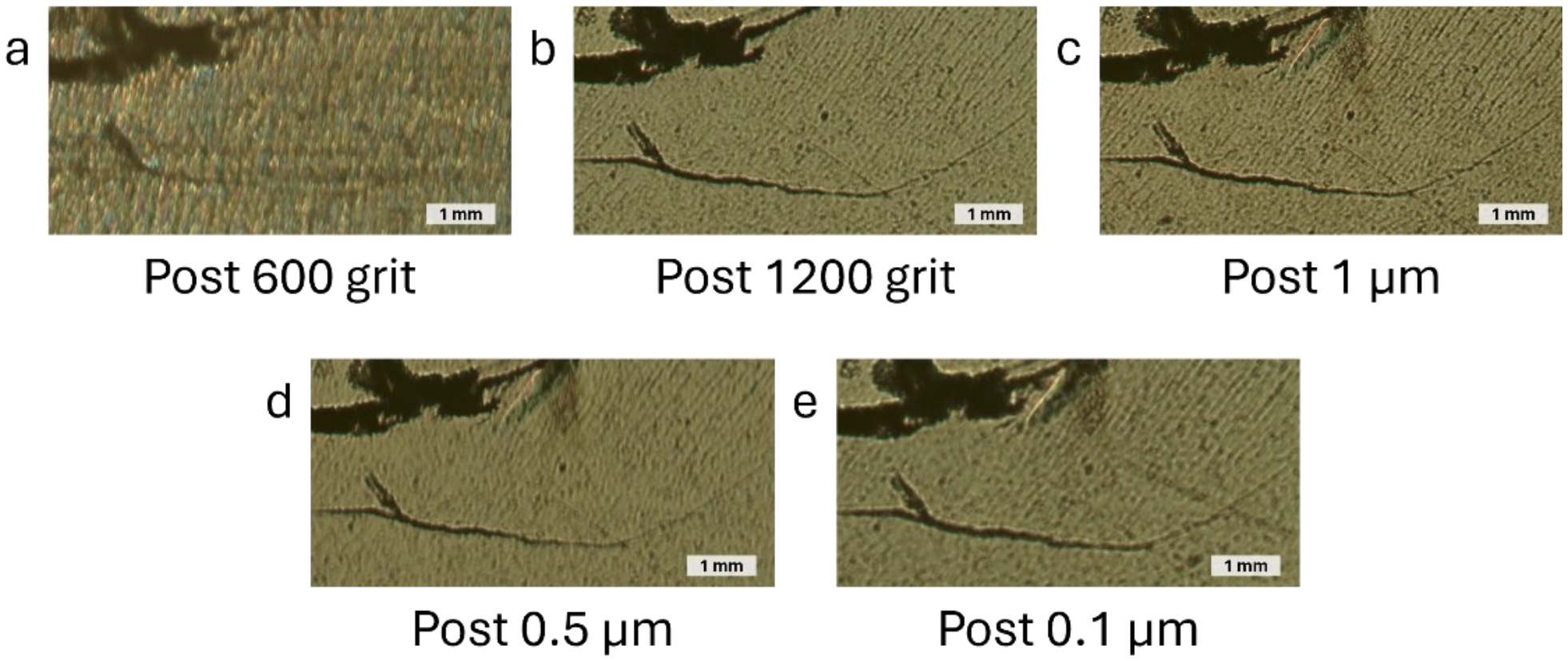

**Figure 5.** Optical images comparing surface finish after polishing steps with (a) 600 grit sandpaper, (b) 1200 grit sandpaper, and (c) 1, (d) 0.5, and (e) 0.1 μm particle size diamond polishing pads. All images are at 5x magnification.

Polishing is a critical first step in surface preparation of a substrate. Optical images of the same region of one sample are presented in Figure 5 between polishing steps (see methods). After polishing with 600 grit sandpaper (Figure 5a) there is visible roughness on the surface of the sample. This roughness becomes less pronounced after polishing with 1200 grit sandpaper (Figure 5b) and was visibly gone after polishing with the 1μm particle size diamond polishing pads (Figure 5c). Little change was observed at this level of magnification upon polishing with finer particle size diamond pads (Figure 5d and Figure 5e).

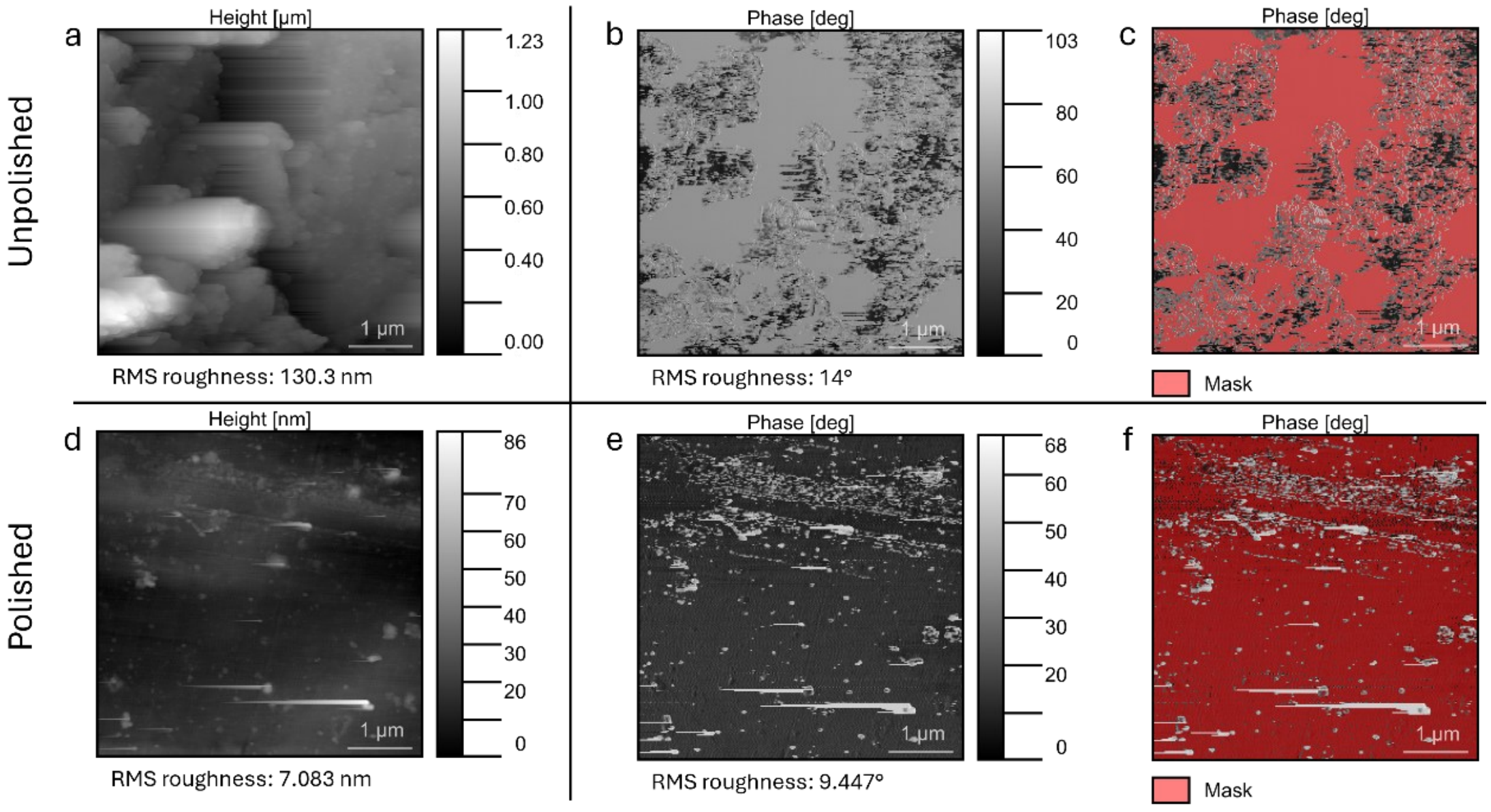

**Figure 6.** Tapping-mode atomic force microscope images of an unpolished (a-c) and polished (d-f) surface of $Ta_{0.8}Hf_{0.2}C_{0.5}$, measuring sample height (a and d) and phase (b, c, e, and f) during tapping. Units of (a) and (d) are different due to the range of heights measured. Masked region of (c) is between 52.44° and 63.74° from the minimum. Masked region of (f) is between 8.52° and 19.07° from the minimum. Values of all images are shifted so the minimum is zero.

Tapping AFM was used to determine how effective the polishing process was in preparing the pieces for use as substrates. A height map and phase map of the data are presented in Figure 6, showing a stark difference between a polished and unpolished piece. All maps are square 5 µm x 5 µm regions, 256 x 256 pixels. Figure 6a shows the variation in height of an unpolished piece. The maximum height is 1.228 µm from the minimum, the root-mean square (RMS) roughness is 130.3 nm, and the mean roughness is 89.9 nm. RMS roughness is a useful metric for determining how flat a surface is and quantifying how significant the variation in surface height is. A RMS roughness value of less than 1 nm is required for epitaxial growth. The phase map of the unpolished surface (Figure 6b) has a maximum phase difference of 102.8° from the minimum, with an RMS roughness of 14°, and a mean roughness of 10.22°. In contrast, Figure 6d shows the height map of the polished surface. The maximum height is 85.05 nm above the minimum, with an RMS roughness of 7.083 nm, and a mean roughness of 5.243 nm. The range on the values of the polished phase map (Figure 6e) is similarly tighter, with the maximum phase difference being 68.12° from minimum, an RMS roughness of 9.477°, and an average roughness of 5.725°. This stark difference in RMS height roughness when comparing a polished and unpolished sample shows that the material responds well to conventional polishing techniques and that further polishing should improve surface planarization. The difference in phase measurements can be attributed to an outer surface layer being removed, making the measured area more homogeneous. Chemo-mechanical polishing (CMP) is shown to further improve the planarization of substrates and remove sub-surface damage induced by mechanical polishing.[40] The demonstration of optically smooth surfaces is a critical first step towards enabling CMP.

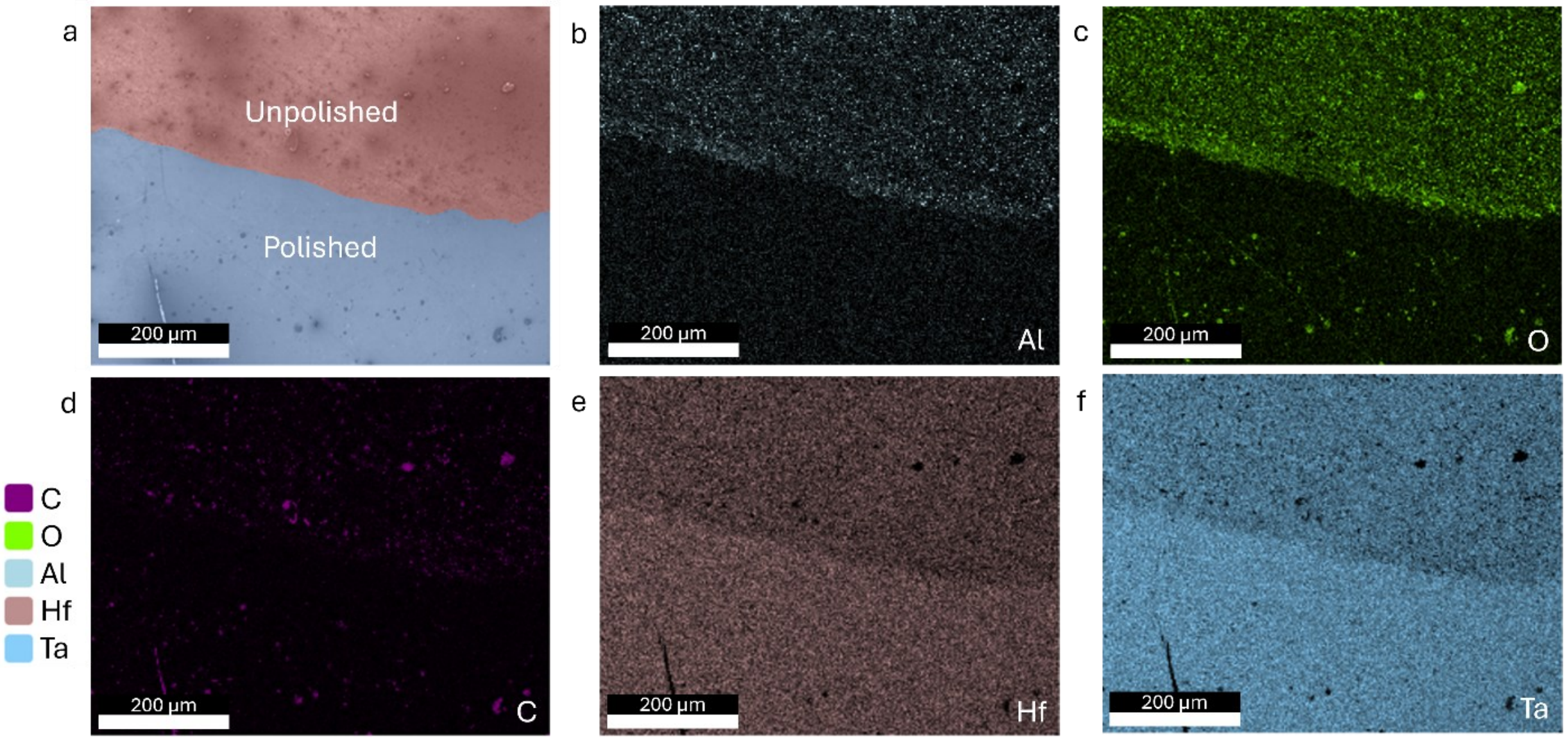

**Figure 7.** Scanning electron microscope images of the interface between a polished and unpolished region. The raw image is overlaid with the polished and unpolished regions (a) after polishing with 0.1 µm particle size diamond

polishing sheets. Individual atomic maps generated by 10 keV radiation are presented for (b) Al, (c) O, (d) C, (e) Hf, and (f) Ta.

Further information on approximate surface chemistry was obtained by SEM. SEM was conducted on the grown rod at the interface of a polished and unpolished region. Figure 7a shows the raw SEM image with the polished region (post 0.1 µm particle size diamond polishing sheets) on the bottom and the region cut with a diamond saw, without any polishing, on the top. Direct comparisons of surface roughness of each can be made as the two surfaces look distinct with the unpolished region being visibly rougher. In Figure 7 a clear compositional gradient is present, with Ta and Hf being more concentrated in the polished region compared to the unpolished region. In the unpolished region, the O content on the surface is significantly higher than on the polished surface as seen in Figure 7c, which is the primary cause of the apparent compositional gradient of the metals. This is due to surface oxides that are present in the unpolished region but were ground away on the mechanically polished surface. C mapping is presented in Figure 7d, showing that C is most concentrated in the scratch present on the surface and the small grains that are more prominent on the rougher side, which is also a large source of O content. The distribution of Ta and Hf within each region is uniform, with the polished surface showing more signal for both than the unpolished surface (Figure 7e and Figure 7f). The exact Hf:Ta ratio cannot be accurately determined, as the Ta M and Hf M peaks overlap significantly (Figure S3). In total, the region scanned reports 46(10)% C, 22(9)% O, 5.4(45)% Hf, and 22.1(48)% Ta by atomic composition. Al seen in Figure 7b is due to contamination on the cutting blade and/or the Al disks used in the mechanical polishing step. The excess of C, concentrated in C-rich inclusions, is attributed to the excess used during arc melting to not be C deficient, as residual grains adhered to the surface from the SiC polishing pads (this is one of the challenges in working with an ultra-hard substrate material).

To better quantify the Hf:Ta ratio, XPS measurements of a cleaned and annealed surface were conducted, shown in Figure 8. The calculated Hf:Ta ratio was 0.12, about half the ~0.24 found by SEM. Quantification of carbon, Figure S4 and S5, gives a metal:C ratio of ~1.9:1, in good agreement with the crystallographic studies showing a 2:1 ratio.

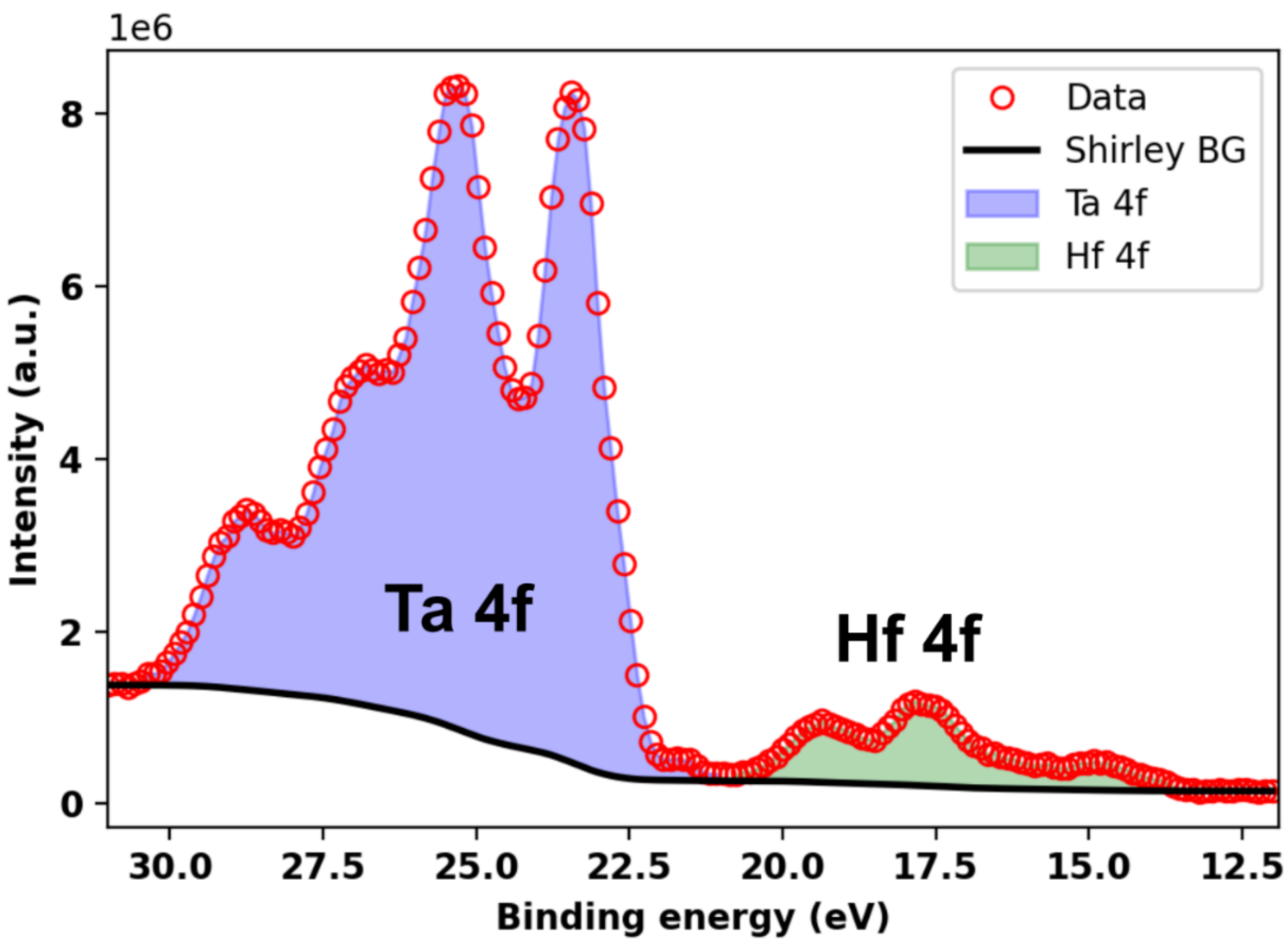

**Figure 8**. XPS of 4f Ta and Hf doublets for a $Ta_{0.8}Hf_{0.2}C_{0.5}$ surface, with highlighted area to extract total Ta:Hf ratio.

Thermal properties of $Ta_{0.8}Hf_{0.2}C_{0.5}$ were measured with TDTR. The ratio of the negative in-phase (-$V_{in}$) and out-of-phase ($V_{out}$) components of the measured lock-in signal was fit with a thermal diffusion model for modulation frequencies of 8.4 MHz, 1.1 MHz and 0.5 MHz (Figure 9).[34] The free parameters used to fit the data to the thermal model included $\kappa$ (cross-plane thermal conductivity), $C$ (volumetric heat capacity) and $G$ (thermal boundary conductance) at the interface of the Al transducer and $Ta_{0.8}Hf_{0.2}C_{0.5}$ substrate. $\kappa$ was tabulated as a function of $C$ for different frequencies as shown in Figure 10.[41] The intersection of these curves provides $\kappa$ and $C$ values of 18.1 W $m^{-1}$ $K^{-1}$ and 2 MJ $m^{-3}$ $K^{-1}$, respectively, for a thermal conductance value of 40 MW $m^{-2}$ $K^{-1}$ at the Al-$Ta_{0.8}Hf_{0.2}C_{0.5}$ interface. The resultant $\kappa$ (18.1 W $m^{-1}$ $K^{-1}$ ) of $Ta_{0.8}Hf_{0.2}C_{0.5}$ measured via TDTR is comparable to the values of TaC (27.9 W $m^{-1}$ $K^{-1}$) and HfC (21W $m^{-1}$ $K^{-1}$), measured using laser flash technique.[42,43] The value of 18.1 W $m^{-1}$ $K^{-1}$ is also comparable to the 34 W $m^{-1}$ $K^{-1}$ found in sapphire,[44] a very common substrate for III-Nitride optoelectronics and lateral power devices,[45] and above that of $Ga_2O_3$ (10.9 W $m^{-1}$ $K^{-1}$),[46] a leading candidate for ultrawide bandgap vertical power electronics. Such vertical devices enable a lower on-resistance (power handling $\propto E_g^6$) compared to lateral devices (power handling $\propto E_g^4$), as codified in the Baliga Figure of Merit,[47] significantly reducing the thermal load from Joule heating in the power device.

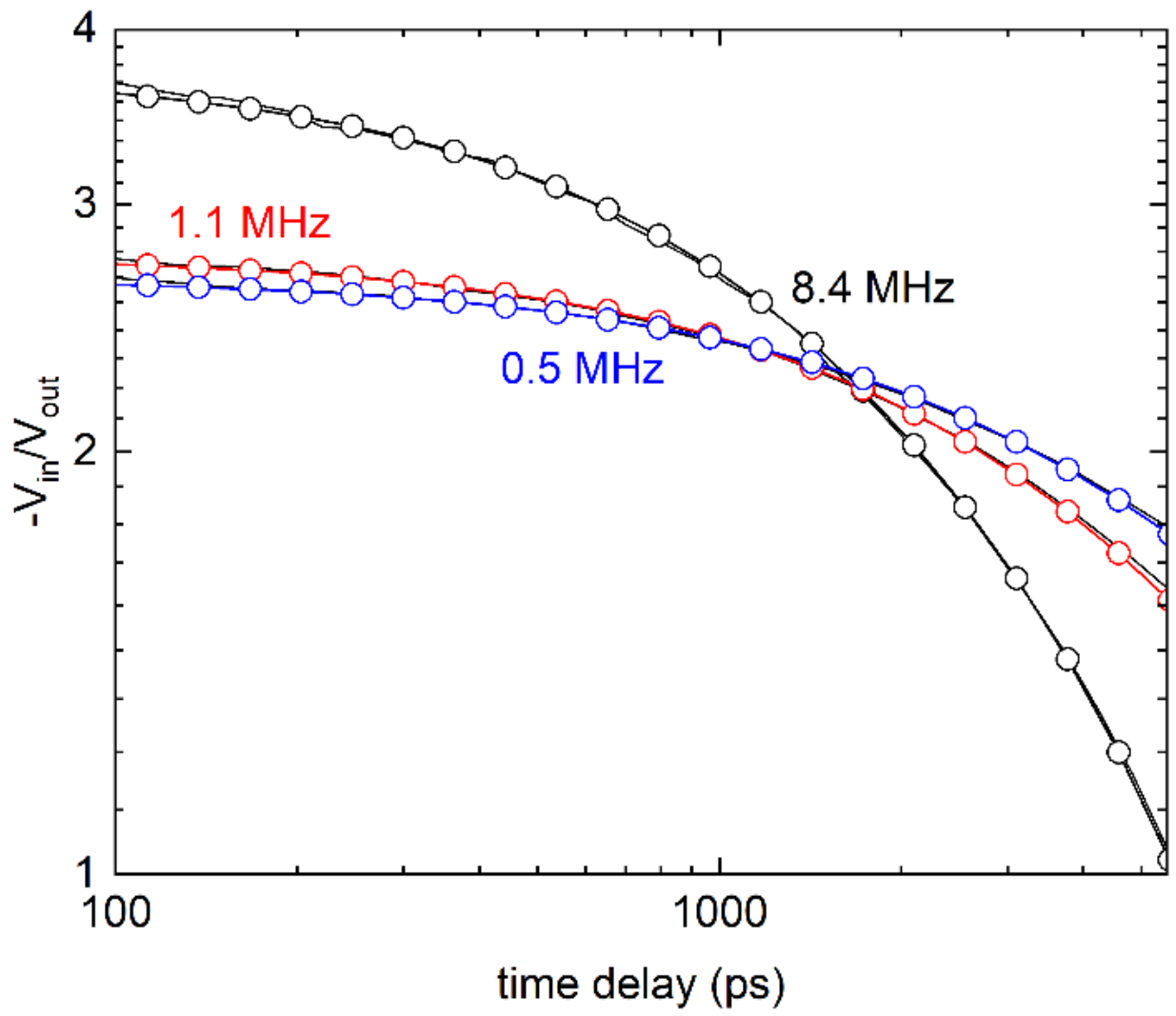

**Figure 9**. TDTR ratio signal, $-V_{in}/V_{out}$ of Al/ $Ta_{0.8}Hf_{0.2}C_{0.5}$ at a frequency of 8.4 MHz, 1.1 MHz and 0.5 MHz with thermal model (unfilled circles) and measured (black solid line) data.

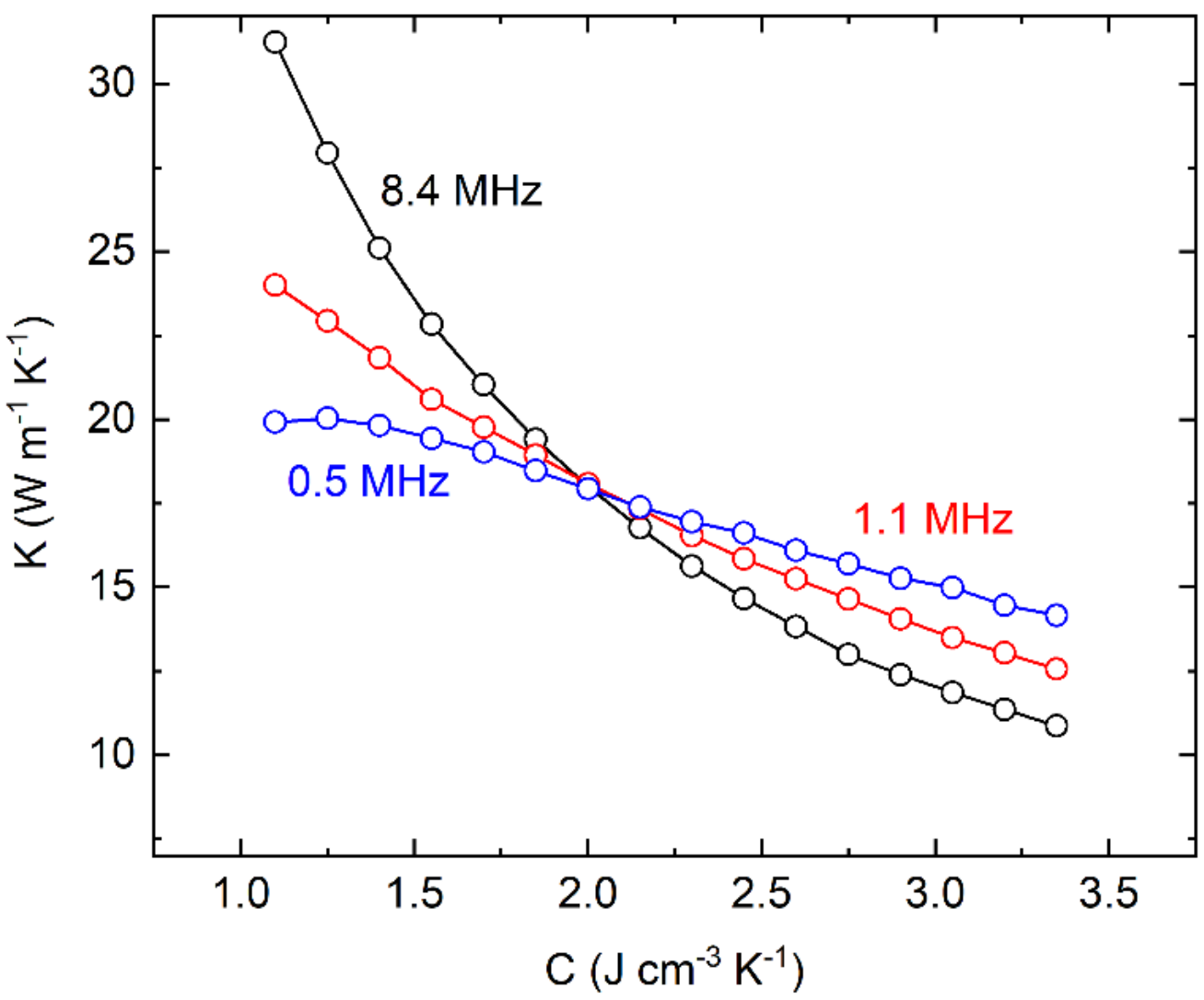

**Figure 10**. *κ* vs *C* plot of $Ta_{0.8}Hf_{0.2}C_{0.5}$ at a frequency of 8.4 MHz, 1.1 MHz and 0.5 MHz measured using frequency-dependent TDTR technique.

No attempts were made in this paper to precisely control the ratio of Ta:Hf, nor to control the combination of $Ta_{1-x}Hf_xC$ and $Ta_{1-x}Hf_xC_{0.5}$, as the material crystallized with the $Ta_{1-x}Hf_xC_{0.5}$ stoichiometry even in the presence of excess C powder. Future efforts should focus in part on controlling the ratio of the metals to understand the impact on the lattice parameters to produce an optimized substrate for the $Al_yGa_{1-y}N$ family.

## Conclusion

Here, we report the successful crystal growth of $Ta_{0.8}Hf_{0.2}C_{0.5}$, and demonstrate its viability as a tunable substrate family for $Al_yGa_{1-y}N$ electronics. Crystal structure solutions via SCXRD have clarified the structure to be trigonal, in line with previous reports of C-deficient $TaC_{1-\delta}$.[48,49] The ratio of Ta:Hf in the polycrystalline disks was found to be near the targeted 4:1, with C segregated in clusters due to the excess C needed to ensure incorporation. AFM demonstrated the efficacy of the polishing procedure on the polycrystalline disks, showing a sharp improvement in RMS roughness while the surface was majority homogeneous. The thermal parameters were found to not differ greatly from the targeted GaN/$Al_yGa_{1-y}N$ films they are intended to enable growth of. Future works lies primarily in utilizing chemo-mechanical planarization (CMP) to make substrates epi-ready and performing growths of $Al_yGa_{1-y}N$ on the cleaned substrates. A more rigorous exploration of the resulting crystal structures found across the $Ta_{1-x}Hf_xC_z$ family, with stronger confirmation of the metal:C ratio, would shed further light on the crystallization pathways of this material family, and its versatility to cover the entire chemical parameter space of $Al_yGa_{1-y}N$.

## Acknowledgements

ENC would like to thank E. Zoghlin for assisting with growth and sample preparation. This work was supported as part of APEX (A Center for Power Electronics Materials and Manufacturing Exploration), an Energy Frontier Research Center funded by the U.S. Department of Energy, Office of Science, Basic Energy Sciences under Award # ERW0345. This research utilized capabilities of the Advanced Photon Source, a U.S. Department of Energy (DOE) Office of Science user facility operated for the DOE Office of Science by Argonne National Laboratory under Contract No. DE-AC02-06CH11357. The National Laboratory of the Rockies (NLR) is operated under Contract No. DE-AC36-08GO28308. This work made use of the bulk crystal growth facility of the National Science Foundation's Platform for the Accelerated Realization, Analysis, and Discovery of Interface Materials (PARADIM) under Cooperative Agreement DMR-2039380. This research used High-Performance Computing resources at NRL. The views expressed in the article do not necessarily represent the views of DOE or the U.S. Government.

# Supporting Information for Crystal growth and characterization of the ultra-high temperature substrate $Ta_{1-x}Hf_xC_{0.5}$

Evan N. Crites[1,2,*], Sharad Mahatara[3], Joshua R. Hummel[4], Sydney R. Laywell[1], Ahamed Raihan[5], Shivashree S. Gowda[6], Ethan A. Scott[6], Amitayush Jha Thakur[7], Jessica L. McChesney[7], Patrick E. Hopkins[6,8,9], MVS Chandrashekhar[5], Michael G. Spencer[5], Stephan Lany[3], Satya K. Kushwaha[1,2], Tyrel M. McQueen[1,2,4,*]

Affiliations:

[1]*Department of Chemistry, The Johns Hopkins University, Baltimore, Maryland 21218, United States*

[2]*William H. Miller III Department of Physics and Astronomy, The Johns Hopkins University, Baltimore, Maryland 21218, United States*

[3]*National Laboratory of the Rockies, Golden, CO 80401, United States*

[4]*Department of Materials Science and Engineering, The Johns Hopkins University, Baltimore, Maryland 21218, United States*

[5]*Center for Research and Education in Microelectronics, Department of Electrical and Computer Engineering, Morgan State University, Baltimore, MD 21251, United States*

[6]*Department of Mechanical and Aerospace Engineering, University of Virginia, Charlottesville, Virginia, 22904, United States*

[7]*Advanced Photon Source, Argonne National Laboratory, Lemont, Illinois 60439, USA*

[8]*Department of Materials Science and Engineering, University of Virginia, Charlottesville, Virginia, 22904, United States*

[9]*Department of Physics, University of Virginia, Charlottesville, Virginia, 22904, United States*

**Table of Contents**

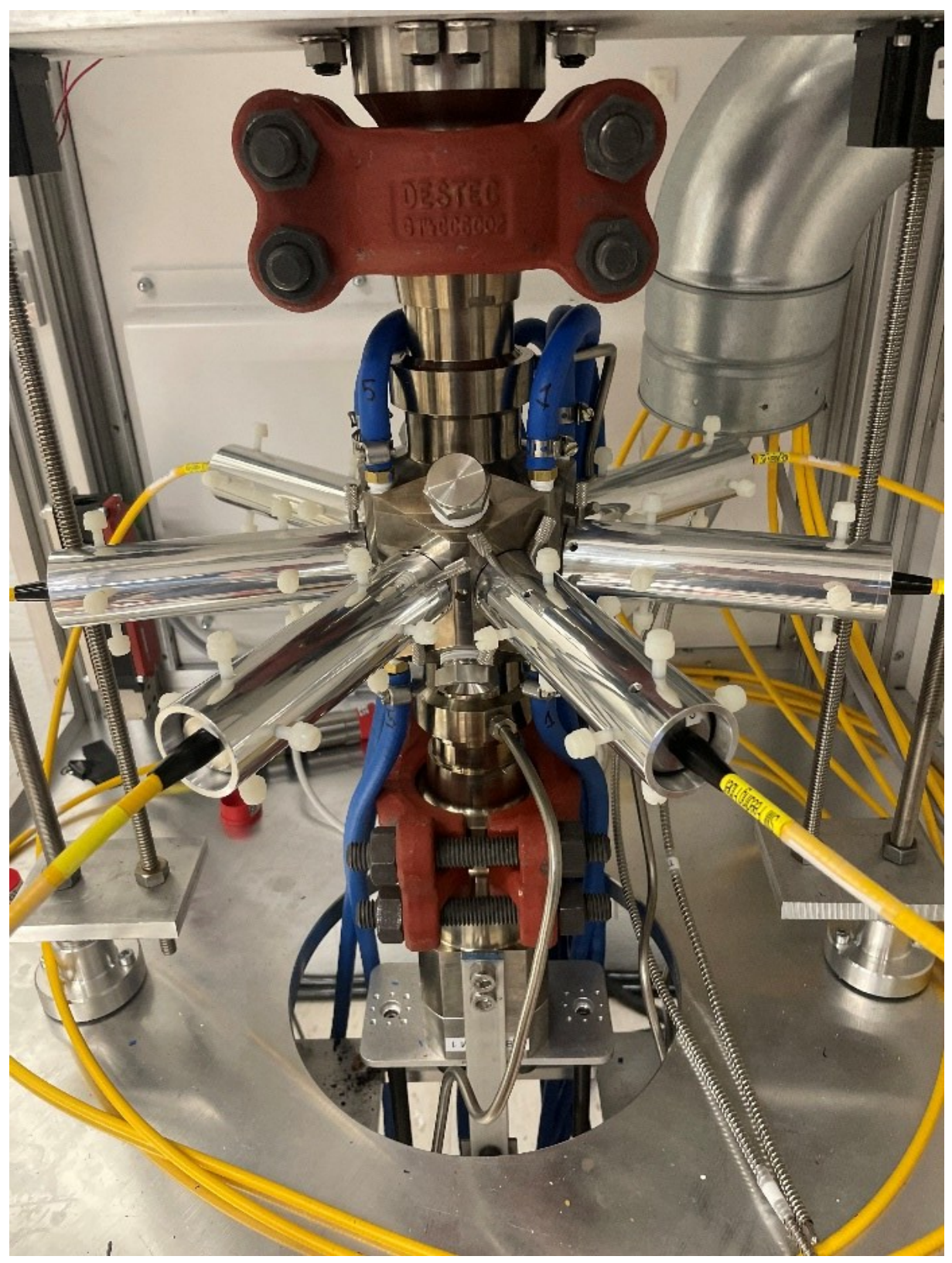

**Figure S1.** Picture of the custom optical floating zone furnace used in this work.

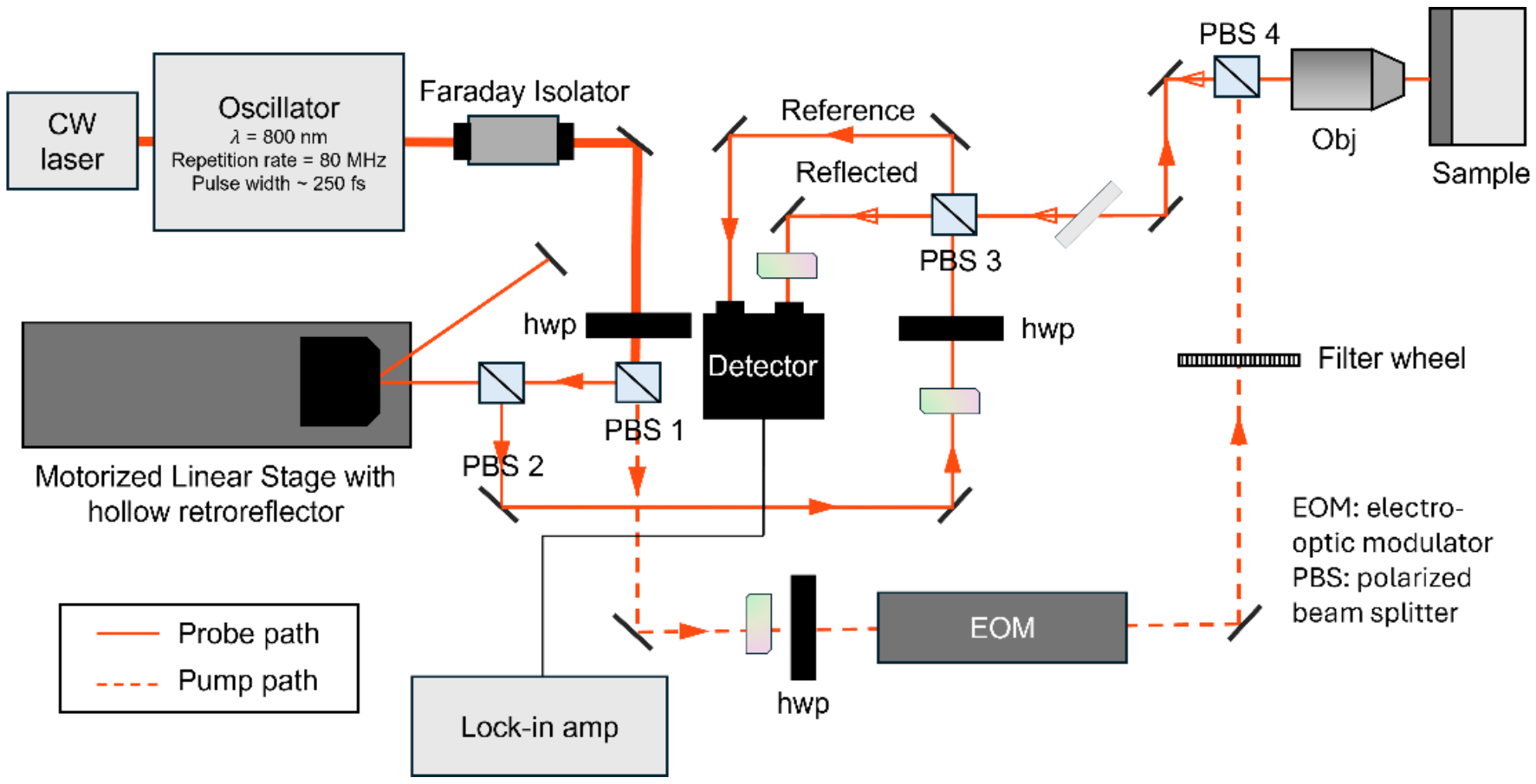

**Figure S2.** Schematic of a two-tint frequency-dependent time-domain thermoreflectance setup.

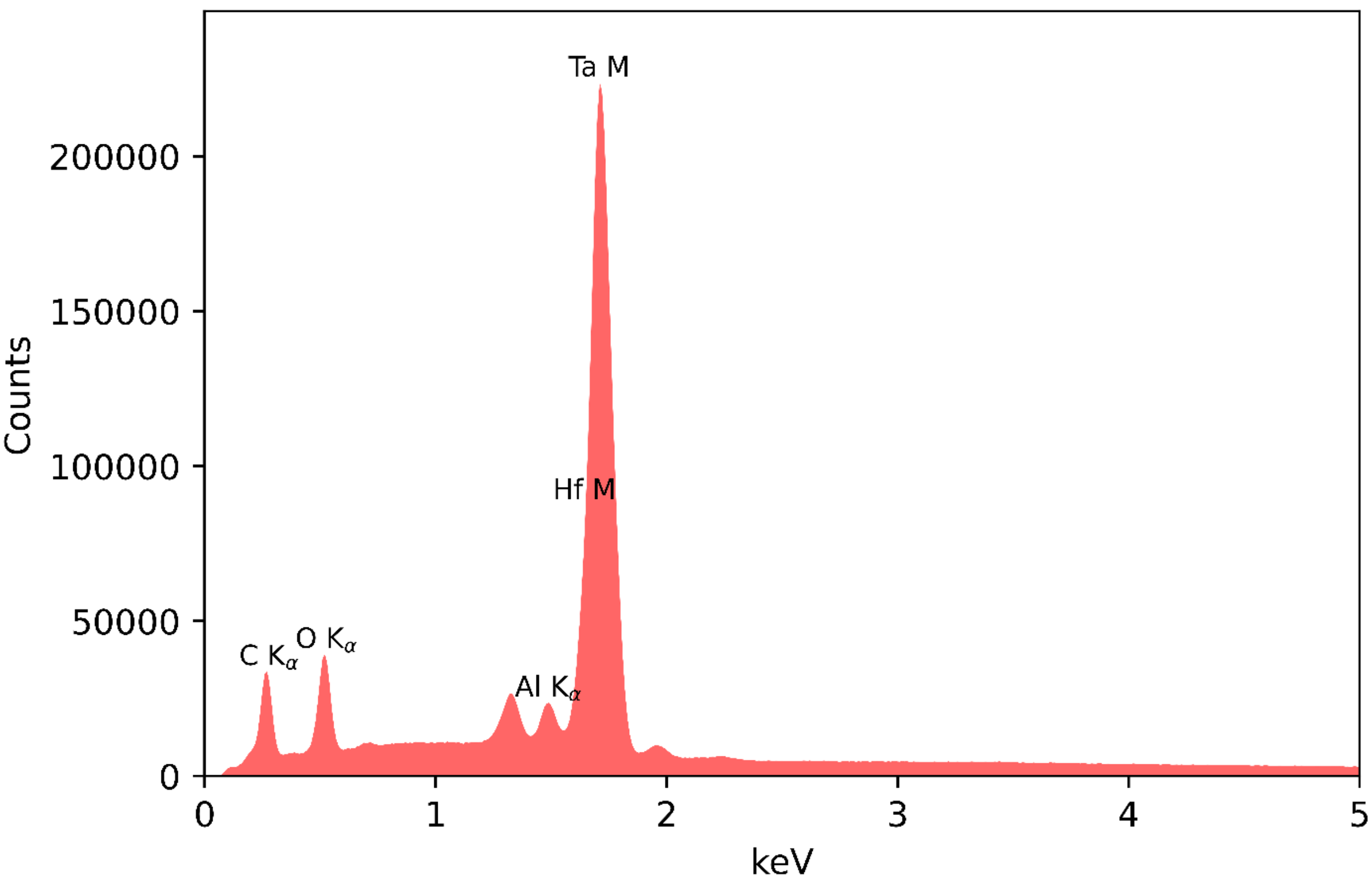

**Figure S3.** EDX spectrum of mapped region. The Hf M and Ta M peaks significantly overlap, making accurate quantification of the Hf:Ta ratio impossible.

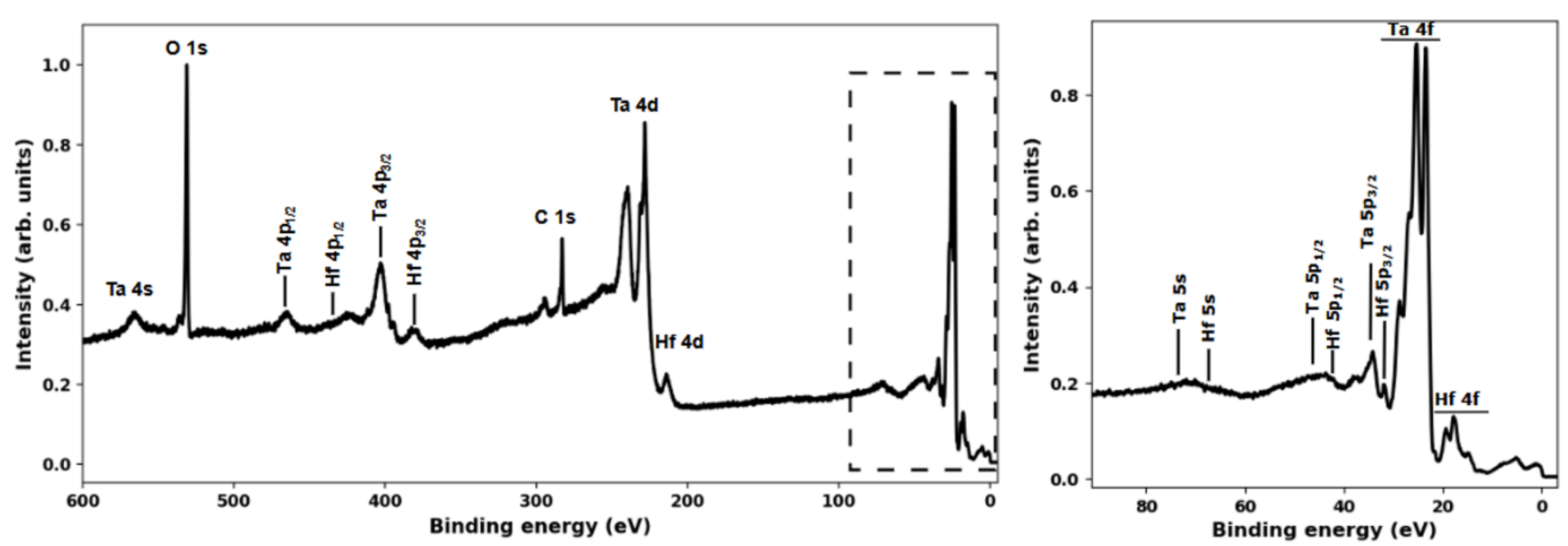

**Figure S4.** Wide energy window XPS spectrum measured using photons at 1486 eV. The region under 100 eV is expanded to show which peaks arise from which elements.

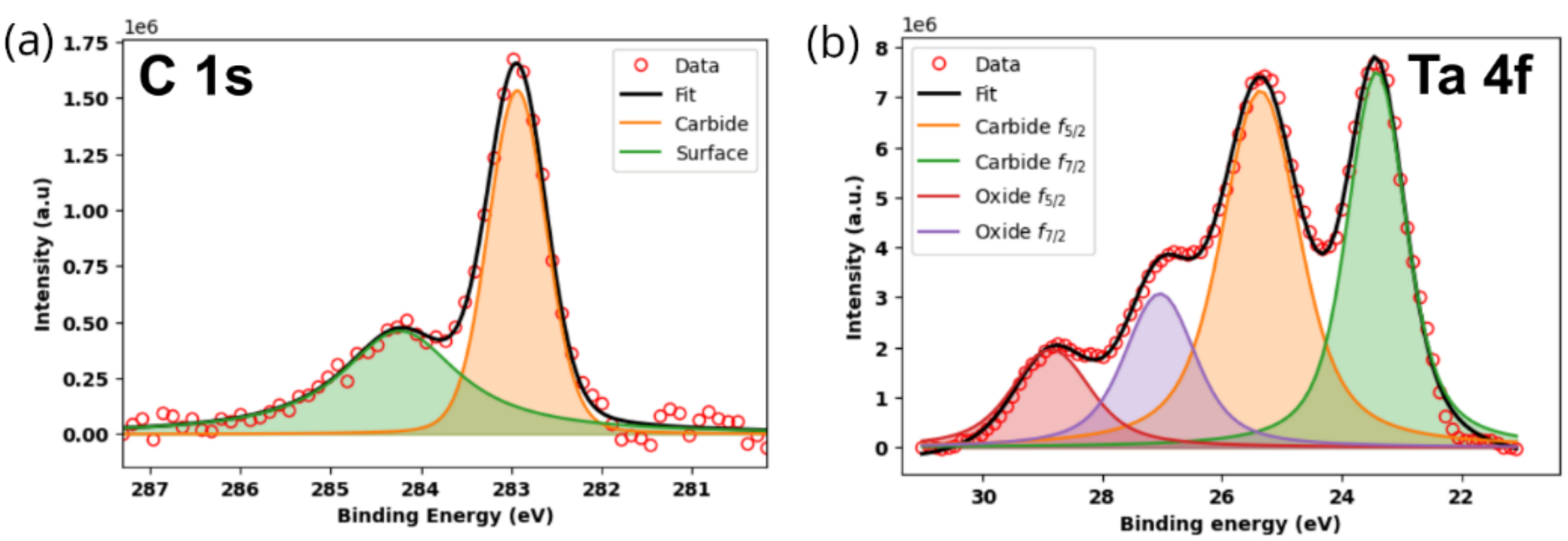

**Figure S5.** XPS spectra of (a) C-1s core level with Voigt peak fitting for carbide and C-C (surface) components, and (b) Ta 4f core levels with Voigt peak fitting for carbide and oxide components. All measurements were done using photons at 1486 eV.